# THE FOURIER TRANSFORM SOLUTION
# for the
# GREEN'S FUNCTION OF MONOENERGETIC NEUTRON TRANSPORT THEORY

B. Ganapol[+]
Department of Aerospace and Mechanical Engineering
University of Arizona

**ABSTRACT**
Nearly 45 years ago, Ken Case published his seminal paper on the singular eigenfunction solution for the Green's function of the monoenergetic neutron transport equation with isotropic scattering. Previously, the solution had been obtained by Fourier transform. While it is apparent the two had to be equivalent, a convincing equivalence proof for general anisotropic scattering remained a challenge until now.

Key Words: Generalized singular eigenfunctions, Fourier transform, anisotropic scattering

**INTRODUCTION**
It has been nearly 45 years since the derivation of the most meaningful analytical solution in neutron transport theory. Ken Case, in a remarkably insightful paper [1]-- one of the most cited papers in all of physics-- applied separation of variables to express the Green's function for monoenergetic, isotropically scattering neutrons in terms of singular eigenfunctions. Mika [2], shortly thereafter, considered Anisotropic scattering. Previously, the isotropic scattering solution had been found through Fourier transform inversion [3,4,5]. Showing equivalence between the singular eigenfunction and Fourier transform solutions for other than isotropic scattering is anything but straightforward (as shown by the author [6]). While equivalence was indeed demonstrated, it was not intuitively obvious and lacked simplicity. Thus, the challenge of showing a straightforward equivalence of the two approaches remained. In the following, we revisit equivalence in a more unified manner through the Legendre polynomial expansion of the Fourier transformed solution in terms of analytically determined moments. In so doing, Chandrasekhar polynomials and generalized singular eigenfunctions naturally emerge to simplify the analytical transform inversion.

---

[+] Ganapol@cowboy.ame.arizona.edu

What follows is not intended to be a more appropriate solution for the Green's function than given by singular eigenfunctions, but merely to represent a different approach to further our knowledge. The motivation follows the reasoning of Case in choosing the method of separation of variables originating from partial differential equations, in that an equivalent solution must be possible through Fourier transforms. The challenge in showing the obvious therefore is in the development of the analytical thinking and corresponding mathematics to meet the objective.

**I. THE MONOENERGETIC GREEN'S FUNCTION**

Our focus is the Green's function for the neutron transport equation,

$$\left[\mu\frac{\partial}{\partial x}+1\right]\psi(x,\mu;\mu_0) = \frac{c}{2}\sum_{l=0}^{\infty}\omega_l P_l(\mu)\psi_l(x;\mu_0)+\delta(\mu-\mu_0)\delta(x), \quad (1a)$$

representing scattering and absorption of neutrons without energy loss in an infinite plane medium satisfying the condition

$$\lim_{|x|\to\infty}\psi(x,\mu;\mu_0)<\infty. \quad (1b)$$

By application of translational invariance, the source, emitting in direction $\mu_0$, is located at $x = 0$ for convenience. The total cross section is unity and $c$ is the number of scattering secondaries $0 \leq c < 1$. $\omega_l$ is the $l^{th}$ scattering coefficient for a Legendre polynomial series expansion of the scattering kernel. At this point, the expansion has an infinite number of terms-- justified if

$$\sum_{l=0}^{\infty}|\omega_l|<\infty. \quad (1c)$$

The Legendre moments,

$$\psi_l(x;\mu_0) \equiv \int_{-1}^{1}d\mu P_l(\mu)\psi(x,\mu;\mu_0);\ l=0,1..., \quad (1d)$$

will play a key role in what follows. As usual, $\mu$ and $x$ are the neutron direction and position of the neutron angular flux distribution $\psi(x,\mu;\mu_0)$.

The approach to show equivalence follows two paths in Fourier transform space, the first of which, while providing a solution, does not readily lead to an explicit analytical form for general anisotropic scattering. The second, through a combination of a moments solution with the first solution, does provide the explicit solution. Eventually, to arrive at the analytical form will require the evaluation of the Fourier transform inversion through contour integration.

## II. STANDARD SOLUTION
Here, a straightforward Fourier inversion gives an image function that, in principle, one can invert.

## II.A. Fourier transform
The standard angular flux solution derives from the Fourier transform of Eq(1a)

$$\bar{\psi}(k,\mu;\mu_0) \equiv \int_{-\infty}^{\infty} dx e^{-ikx} \psi(x,\mu;\mu_0), \qquad (2a)$$

and the Legendre moments

$$\bar{\psi}_l(k;\mu_0) \equiv \int_{-\infty}^{\infty} dk e^{-ikx} \psi_l(x;\mu_0), \qquad (2b)$$

to give

$$(1+ik\mu)\bar{\psi}(k,\mu;\mu_0) = \frac{c}{2}\sum_{l=0}^{\infty} \omega_l P_l(\mu)\bar{\psi}_l(k;\mu_0) + \delta(\mu-\mu_0). \qquad (3)$$

Solving for the angular flux in terms of moments yields

$$\bar{\psi}(k,\mu;\mu_0) =$$
$$= \frac{c}{2}\frac{1}{(1+ik\mu)}\sum_{l=0}^{\infty} \omega_l P_l(\mu)\bar{\psi}_l(k;\mu_0) + \frac{1}{(1+ik\mu)}\delta(\mu-\mu_0), \qquad (4a)$$

from which the moments themselves are found by projection over Legendre polynomials

$$\bar{\psi}_j(k;\mu_0) = c\sum_{l=0}^{\infty} \omega_l L_{j,l}(z)\bar{\psi}_l(k;\mu_0) + \frac{z}{z+\mu_0} P_j(\mu_0); j = 0,1,.... , \quad (4b)$$

where the matrix element is explicit

$$L_{j,l}(z) \equiv \frac{z}{2}\int_{-1}^{1} d\mu \left[\frac{P_j(\mu)P_l(\mu)}{z+\mu}\right]$$

$$= (-1)^{l+j} z \begin{cases} Q_l(z)P_j(z), & j \leq l \\ Q_j(z)P_l(z), & l < j. \end{cases} \quad (4c)$$

In the above, $z \equiv 1/ik$.

To solve Eq(4b) requires truncation of the scattering kernel, where we assume the scattering coefficient $\omega_l$ to vanish for $l \geq L+1$ to give the linear set of equations

$$\sum_{l=0}^{L} \left[\delta_{j,l} - c\omega_l L_{j,l}(z)\right]\bar{\psi}_l(k;\mu_0) = \frac{z}{z+\mu_0} P_j(\mu_0), \quad (5a)$$

or in vector notation

$$[\mathbf{I} - c\mathbf{L}(z)\mathbf{W}]\bar{\boldsymbol{\psi}}(k;\mu_0) = \frac{z}{z+\mu_0}\mathbf{P}(\mu_0), \quad (5b)$$

with

$$\mathbf{L}(z) \equiv \{L_{j,l}(z); j,l = 1,2,...,L\}$$
$$\mathbf{W} \equiv diag(\omega_l)$$
$$\mathbf{P}(\mu_0) \equiv \{P_j(\mu_0); j = 0,...,L\}.$$

The solution to Eq (5b) by matrix inversion,

$$\bar{\boldsymbol{\psi}}(k;\mu_0) = \frac{z}{z+\mu_0}\left[\boldsymbol{I} - c\boldsymbol{L}(z)\boldsymbol{W}\right]^{-1}\boldsymbol{P}(\mu_0), \qquad (6)$$

gives the angular moments vector,

$$\bar{\boldsymbol{\psi}}(k;\mu_0) = \left[\bar{\psi}_1(k;\mu_0) \quad \bar{\psi}_2(k;\mu_0) \quad \ldots \quad \bar{\psi}_L(k;\mu_0)\right]^T.$$

**II.B. Fourier transform inversion**
The angular flux transform comes from Eq(4a)

$$\bar{\psi}(k,\mu;\mu_0) = \frac{z}{z+\mu_0}\delta(\mu-\mu_0) + \frac{c}{2}\frac{z}{z+\mu}\boldsymbol{P}^T(\mu)\boldsymbol{W}\bar{\boldsymbol{\psi}}(k;\mu_0). \qquad (7)$$

From the inversion integral

$$\psi(x,\mu;\mu_0) \equiv \frac{1}{2\pi}\int_{-\infty}^{\infty} dk\, e^{ikx}\bar{\psi}(k,\mu;\mu_0), \qquad (8)$$

there results

$$\psi(x,\mu;\mu_0) = \frac{e^{-|x/\mu_0|}}{|\mu_0|}\Theta(x/\mu_0)\delta(\mu-\mu_0) + \\ + \frac{c}{4\pi}\boldsymbol{P}^T(\mu)\boldsymbol{W}\int_{-\infty}^{\infty} dk\, e^{ikx}\frac{z}{z+\mu}\frac{z}{z+\mu_0}\left[\boldsymbol{I} - c\boldsymbol{L}(z)\boldsymbol{W}\right]^{-1}\boldsymbol{P}(\mu_0), \qquad (9)$$

where $\Theta(x/\mu_0)$ is the unit step function in the uncollided contribution.

While Eq(9) is a valid solution, it involves matrix inversion inside the Fourier inversion integral and, unless performed analytically, obscures any analytical form of contour integration. One can only say poles come from the zeros of *Det* $\left[\boldsymbol{I} - c\boldsymbol{L}(z)\boldsymbol{W}\right]$ and branch points $\pm i$ come from the logarithm in the Legendre functions in $\boldsymbol{L}$; otherwise, Eq(10) is more suited to numerical rather than analytical investigation [7].

## III. NON-STANDARD SOLUTION

Since the analytical inversion of Eq(9) is not sufficiently explicit, we seek a more informative one.

### III.A. General solution to the moments recurrence

We find a more useful solution by projection of Eq(3) over Legendre polynomials to give the following three–term recurrence for the Legendre moments:

$$zh_l\bar{\psi}_l(k;\mu_0) + (l+1)\bar{\psi}_{l+1}(k;\mu_0) + l\bar{\psi}_{l-1}(k;\mu_0) = zS_l(\mu_0), \qquad (10)$$

with

$$h_l \equiv 2l + 1 - \omega_l$$
$$S_l(\mu_0) \equiv (2l+1)P_l(\mu_0).$$

From the theory of recurrences [8], the general solution for the moments is

$$\bar{\psi}_l(k;\mu_0) = a(z;\mu_0)g_l(z) + b(z;\mu_0)\rho_l(z) + z\sum_{j=0}^{l}\alpha_{l,j}(z)S_j(\mu_0), \qquad (11a)$$

where the functions $g_l(z)$ and $\rho_l(z)$ are solutions to the homogeneous from of Eq(10); followed by the particular solution. The coefficients, $a(z;\mu_0)$ and $b(z;\mu_0)$, are to be determined from the starting condition at $l = 0$,

$$\bar{\psi}_0(k;\mu_0) = a(z;\mu_0)g_0(z) + b(z;\mu_0)\rho_0(z) + z\alpha_{0,0}(z)S_0(\mu_0),$$

most simply met if

$$a(z;\mu_0) \equiv \bar{\psi}_0(k;\mu_0);\ g_0(z) \equiv 1;\ \rho_0(z) \equiv 0;\ \alpha_{0,0}(z) \equiv 0. \qquad (11b)$$

On substitution of Eqs(11) into the recurrence, there results

$$a(z;\mu_0)\left[zh_l g_l(z)+(l+1)g_{l+1}(z)+lg_{l-1}(z)\right]+$$
$$+b(z;\mu_0)\left[zh_l \rho_l(z)+(l+1)\rho_{l+1}(z)+l\rho_{l-1}(z)\right]+$$
$$+(l+1)\alpha_{l+1,l+1}(z)zS_{l+1}(\mu_0)-l\alpha_{l-1,l}(z)zS_l(\mu_0)+ \qquad (12)$$
$$+z\sum_{j=0}^{l}\left[\begin{array}{c}zh_l\alpha_{l,j}(z)+(l+1)\alpha_{l+1,j}(z)+\\ +l\alpha_{l-1,j}(z)\end{array}\right]S_j(\mu_0)=zS_l(\mu_0).$$

We now consider two cases. The first is for $l > 0$, and requires the terms in square brackets in Eq(12) to vanish, i.e.,

$$zh_l g_l(z)+(l+1)g_{l+1}(z)+lg_{l-1}(z)\equiv 0$$
$$zh_l \rho_l(z)+(l+1)\rho_{l+1}(z)+l\rho_{l-1}(z)\equiv 0$$
$$zh_l \alpha_{l,j}(z)+(l+1)\alpha_{l+1,j}(z)+l\alpha_{l-1,j}(z)=0$$

leaving

$$(l+1)\alpha_{l+1,l+1}(z)zS_{l+1}(\mu_0)-l\alpha_{l-1,l}zS_l(\mu_0)=zS_l(\mu_0),$$

from which

$$l\alpha_{l-1,l}(z)\equiv -1$$
$$\alpha_{l+1,l+1}(z)\equiv 0$$

(13)

readily follows. The second case is for $l = 0$

$$b(z;\mu_0)\left[zh_l \rho_l(z)+(l+1)\rho_{l+1}(z)+l\rho_{l-1}(z)\right]_{l=0}+$$
$$+\alpha_{1,1}(z)zS_1(\mu_0)=zS_0(\mu_0).$$

With $\rho_{-1}$ and $\alpha_{-1,0}$ arbitrarily set to zero and since $\rho_0$ is already zero

$$b(z;\mu_0)\rho_1(z)+\alpha_{1,1}(z)zS_1(\mu_0)=zS_0(\mu_0)$$

suggesting

$$\rho_1(z) \equiv -z,$$
$$b(z;\mu_0) \equiv -S_0(\mu_0) = -1$$
$$\alpha_{1,1}(z) \equiv 0,$$

where $\rho_1(z)$ has been chosen to conform to conventional normalization of the Chandrasekhar polynomial of the second kind to be defined below.

To summarize, for $l \geq 0$, we have

$$g_0(z) = 1$$
$$zh_l g_l(z) + (l+1) g_{l+1}(z) + l g_{l-1}(z) = 0 \tag{14a}$$

$$\rho_0(z) = 0$$
$$\rho_1(z) = -z$$
$$zh_l \rho_l(z) + (l+1) \rho_{l+1}(z) + l \rho_{l-1}(z) = 0 \tag{14b}$$

$$\alpha_{l,l}(z) = 0$$
$$zh_l \alpha_{l,j}(z) + (l+1) \alpha_{l+1,j}(z) + l \alpha_{l-1,j}(z) = 0. \tag{14c}$$

In addition, quantities with negative subscripts vanish. The first two recurrences will ultimately lead to Chandrasekhar polynomials.

For the solution to the recurrence of Eqs(14c), we note that for each $j$, the recurrence satisfies the same as those for $g_l$ and $\rho_l$ for $l \geq 1$ leading to the most general solution

$$\alpha_{l,j}(z) = u_j(z) g_l(z) + v_j(z) \rho_l(z). \tag{15}$$

Using Eq(13) then gives

$$\alpha_{l,l}(z) = u_l(z)g_l(z) + v_l(z)\rho_l(z) = 0$$

$$\alpha_{l-1,l}(z) = u_l(z)g_{l-1}(z) + v_l(z)\rho_{l-1}(z) = -\frac{1}{l},$$

and solving for $u_l(z)$ and $v_l(z)$

$$u_l(z) = \frac{\rho_l(z)}{l[g_l(z)\rho_{l-1}(z) - g_{l-1}(z)\rho_l(z)]}$$

$$v_l(z) = -\frac{g_l(z)}{l[g_l(z)\rho_{l-1}(z) - g_{l-1}(z)\rho_l(z)]}.$$

(16a,b)

Further simplification of this expression comes from multiplying Eq(14a) by $\rho_l$ and Eq(14b) by $g_l$ and subtracting to give

$$t_{l+1}(z) - t_l(z) = 0,$$

where

$$t_l(z) \equiv l[g_l(z)\rho_{l-1}(z) - g_{l-1}(z)\rho_l(z)].$$

On noting $t_1(z) = z$ and since $t_{l+1}(z) = t_l(z) = t_1(z)$, one finds for the denominator of Eqs(16a,b)

$$l[g_l(z)\rho_{l-1}(z) - \rho_l(z)g_{l-1}(z)] = z, \qquad (17)$$

and Eq(15a) becomes

$$z\alpha_{l,j}(z) = \rho_j(z)g_l(z) - g_j(z)\rho_l(z). \qquad (18)$$

Equation (17) is known as the Liouville- Ostrogradski (L-O) formula and is analogous to the Wronskian for a second order ODE.

Hence, Eq(11a), the solution to the recurrence, becomes

$$\bar{\psi}_l(k;\mu_0) = g_l(z)\bar{\psi}_0(k;\mu_0) - \rho_l(z) +$$
$$+ z\alpha_{l,0}(z) + \sum_{j=1}^{l}\left[\rho_j(z)g_l(z) - g_j(z)\rho_l(z)\right]S_j(\mu_0).$$

Since

$$\alpha_{0,0}(z) \equiv 0$$
$$zh_l\alpha_{l,0}(z) + (l+1)\alpha_{l+1,0}(z) + l\alpha_{l-1,0}(z) \equiv 0,$$

we see that $\alpha_{l,0}(z) \equiv 0$, and the solution to the recurrence becomes

$$\bar{\psi}_l(k;\mu_0) = g_l(z)\bar{\psi}_0(k;\mu_0) -$$
$$- \sum_{j=0}^{l}(2j+1)\left[g_j(z)\rho_l(z) - \rho_j(z)g_l(z)\right]P_j(\mu_0). \tag{19}$$

To conform to conventional definitions of Chandrasekhar polynomials, let

$$g_l(z) \to (-1)^l g_l(z)$$
$$\rho_l(z) \to (-1)^l \rho_l(z)$$

to give for Eq(19)

$$\bar{\psi}_l(k;\mu_0) = g_l(-z)\bar{\psi}_0(k;\mu_0) - \chi_l(-z,\mu_0), \tag{20a}$$

with $\chi_l$

$$\chi_l(z,\mu) \equiv \sum_{j=0}^{l}(2j+1)P_j(\mu)\left[\rho_l(z)g_j(z) - g_l(z)\rho_j(z)\right]. \tag{20b}$$

$g_l(z)$ and $\rho_l(z)$ are now Chandrasekhar polynomials of first and second kinds respectively satisfying the following recurrences:

$$g_l(z), \rho_l(z) = 0; \; l < 0$$

$$g_0(z) = 1$$

$$\rho_0(z) = z$$

$$zh_l \begin{bmatrix} g_l(z) \\ \rho_l(z) \end{bmatrix} - (l+1) \begin{bmatrix} g_{l+1}(z) \\ \rho_{l+1}(z) \end{bmatrix} - l \begin{bmatrix} g_{l-1}(z) \\ \rho_{l-1}(z) \end{bmatrix} = \begin{bmatrix} 0 \\ 0 \end{bmatrix}; \; l \geq 1. \qquad (21\text{a,b})$$

Moreover, the L-O formula [Eq(17)] becomes

$$l\left[\rho_l(z) g_{l-1}(z) - g_l(z) \rho_{l-1}(z)\right] = z. \qquad (22)$$

At this point, we lack the (scalar) moment $\bar{\psi}_0(k; \mu_0)$ to complete the solution.

**A.1 Transport closure**

To continue, we need $\bar{\psi}_0(k; \mu_0)$ or, in other words, closure of the moments equations [Eq(10)]. We apply the exact "transport closure" from Eq(4b) with $j$ set to zero

$$\sum_{l=0}^{L} \left[\delta_{0,l} - c\omega_l L_{0,l}(z)\right] \bar{\psi}_l(k; \mu_0) = \frac{z}{z + \mu_0},$$

to give on substitution

$$\bar{\psi}_0(k; \mu_0) = \frac{1}{\Lambda_L(z)} \left[\frac{z}{z + \mu_0} - cz \sum_{l=0}^{L} \omega_l Q_l(-z) \chi_l(-z, \mu_0)\right] \qquad (23\text{a})$$

with

$$\Lambda_L(z) \equiv 1 - cz \sum_{l=0}^{L} \omega_l Q_l(z) g_l(z). \qquad (23\text{b})$$

The last expression is recognized as the dispersion relation for monoenergetic neutron transport [9] and is an even function.

The dispersion relation is also

$$\Lambda_L(z) = (L+1)\left[g_{L+1}(z)Q_L(z) - g_L(z)Q_{L+1}(z)\right] \qquad (23c)$$

shown as follows. We find an alternative expression by considering the recurrences for $g_l(z)$ and $Q_l(z)$ for $l \geq 1$,

$$zh_l g_l(z) - (l+1)g_{l+1}(z) - lg_{l-1}(z) = 0$$
$$z(2l+1)Q_l(z) - (l+1)Q_{l+1}(z) - lQ_{l-1}(z) = 0.$$

Multiplying the first by $Q_l(z)$ and the second by $g_l(z)$ and subtracting gives

$$cz\omega_l Q_l(z)g_l(z) + t_{l+1} - t_l = 0,$$

where

$$t_l \equiv g_l(z)Q_{l-1}(z) - g_{l-1}(z)Q_l(z).$$

On summation

$$t_{L+1} = t_1 - z\sum_{l=1}^{L} \omega_l Q_l(z) g_l(z),$$

and since

$$t_1 = g_1(z)Q_0(z) - g_0(z)Q_1(z) = 1 - czQ_0(z),$$

Eq(23c) follows. Manipulating recurrences in this way is responsible for the first three expressions in Appendix A1 and virtually all expressions in Appendix A2. Appendices A1 and A2 are included to be a ready reference for the derivations to follow.

We leave any further simplification of $\bar{\psi}_0(k;\mu_0)$ until the establishment of the Fourier transform for the angular flux. Using the moments transform, we now consider the derivation of angular flux transform.

### III.B. The Fourier transform of the angular flux

We begin with the angular flux transform expressed as the Legendre polynomial expansion

$$\psi_L(k,\mu;\mu_0) = \sum_{l=0}^{\infty} \frac{(2l+1)}{2} \bar{\psi}_{l,L}(k;\mu_0) P_l(\mu), \tag{24}$$

where the moments $\bar{\psi}_{l,L}(k;\mu_0)$ are from Eqs(20). The subscript $L$ is added to indicate truncated scattering. When the moments are introduced into Eq(24) with $k$ replaced by $-k$ (for notational convenience), there results

$$\psi_L(-k,\mu;\mu_0) = \phi(z,\mu)\bar{\psi}_{0,L}(-k;\mu_0) - T(z,\mu;\mu_0) \tag{25a}$$

with

$$\phi(z,\mu) \equiv \sum_{l=0}^{\infty} \frac{(2l+1)}{2} g_l(z) P_l(\mu) \tag{25b}$$

and

$$T(z,\mu;\mu_0) \equiv \sum_{l=0}^{\infty} \frac{(2l+1)}{2} \chi_l(z,\mu_0) P_l(\mu). \tag{25c}$$

The sum in Eq(25b) requires further clarification whereby a generalized singular eigenfunction will result.

### B.1 Emergence of generalized singular eigenfunctions

To proceed, consider the limit

$$\phi(z,\mu) = \lim_{N\to\infty} \sum_{l=0}^{N} \frac{(2l+1)}{2} g_l(z) P_l(\mu), \tag{26a}$$

along with the Christoffel-Darboux formula, Eq(A2.9), as derived from recurrences for $g_l(z)$ and $P_l(z)$ (similar to the above procedure) in Ref. [10]

$$\sum_{l=0}^{N}\frac{(2l+1)}{2}g_l(z)P_l(\mu)=\frac{1}{z-\mu}\left\{(N+1)\left[g_{N+1}(z)P_N(\mu)-g_N(z)P_{N+1}(\mu)\right]+czg_L^*(z,\mu)\right\}$$

(26b)

where

$$g_l^*(z,\mu)\equiv\begin{cases}\sum_{j=0}^{l}\omega_j g_j(z)P_j(\mu),\ l\le L\\ g_L^*(z,\mu),\ l\ge L+1.\end{cases}$$

Thus, for $N\ge L+1$,

$$\phi(z,\mu)=\frac{cz}{2}\frac{g_L^*(z,\mu)}{z-\mu}+\lim_{N\to\infty}s_N(z,\mu),\qquad(27a)$$

with

$$s_N(z,\mu)\equiv\frac{(N+1)}{2}\left[\frac{g_{N+1}(z)P_N(\mu)-g_N(z)P_{N+1}(\mu)}{z-\mu}\right].\qquad(27b)$$

The Chandrasekhar polynomials $g_N(z)$ and $g_{N+1}(z)$ for $N\ge L+1$ in the last expression take on an especially simple form as now shown. Since $\omega_l\equiv 0$ for $l\ge L+1$, the recurrence for $g_l(z)$ becomes

$$z(2l+1)g_l(z)-(l+1)g_{l+1}(z)-lg_{l-1}(z)\equiv 0,$$

and is satisfied by Legendre polynomials and functions, as two independent solutions, giving the general solution

$$g_l(z)=a(z)P_l(z)+b(z)Q_l(z).$$

But $g_L(z)$ and $g_{L+1}(z)$ are known from the standard recurrence of Eq(21a); therefore, the unknown coefficients $a(z)$ and $b(z)$ come from letting $l = L$ and $L+1$ in the last equation and solving to give

$$g_l(z) = \Lambda_L(z) P_l(z) + \psi_L(z) Q_l(z), \tag{28a}$$

where

$$\psi_l(z) = (l+1)\left[g_l(z) P_{l+1}(z) - g_{l+1}(z) P_l(z)\right]. \tag{28b}$$

Similarly, for the Chandrasekhar polynomial of the second kind,

$$\rho_l(z) = \gamma_L(z) P_l(z) + \theta_L(z) Q_l(z), \tag{29a}$$

where

$$\gamma_l(z) \equiv (l+1)\left[\rho_{l+1}(z) Q_l(z) - \rho_l(z) Q_{l+1}(z)\right] \tag{29b}$$

$$\theta_l(z) = (l+1)\left[\rho_l(z) P_{l+1}(z) - \rho_{l+1}(z) P_l(z)\right]. \tag{29c}$$

With $l = N \geq L+1$, $g_N(z)$ of Eq(28a) introduced into Eq(27b) gives

$$s_N(z, \mu) = \delta_N(z, \mu) \Lambda_L(z) + q_N(z, \mu) \psi_L(z), \tag{30a}$$

with

$$\delta_N(z, \mu) \equiv \frac{(N+1)}{2}\left[\frac{P_{N+1}(z) P_N(\mu) - P_N(z) P_{N+1}(\mu)}{z - \mu}\right] \tag{30b}$$

and

$$q_N(z, \mu) \equiv \frac{(N+1)}{2}\left[\frac{Q_{N+1}(z) P_N(\mu) - Q_N(z) P_{N+1}(\mu)}{z - \mu}\right]. \tag{30c}$$

Since, it is well known from the Christoffel- Darboux theorem [11] that

$$\sum_{l=0}^{N}\frac{(2l+1)}{2}P_l(z)P_l(\mu) = \frac{N+1}{2}\left[\frac{P_{N+1}(z)P_N(\mu)-P_N(z)P_{N+1}(\mu)}{z-\mu}\right]$$

and from the formal Legendre polynomial series representation of $\delta(z-\mu)$ for $|\mu|\le 1$

$$\sum_{l=0}^{\infty}\frac{(2l+1)}{2}P_l(z)P_l(\mu) = \delta(z-\mu), \qquad (31a)$$

it follows that

$$\lim_{N\to\infty}\left\{\frac{N+1}{2}\left[\frac{P_{N+1}(z)P_N(\mu)-P_N(z)P_{N+1}(\mu)}{z-\mu}\right]\right\} = \delta(z-\mu). \qquad (31b)$$

Similarly, from Eq(A1.2)

$$\lim_{N\to\infty}\left\{\frac{N+1}{2}\left[\frac{Q_N(z)P_{N+1}(\mu)-Q_{N+1}(z)P_N(\mu)}{z-\mu}\right]\right\} = 0. \qquad (31c)$$

Thus, Eq(27a) becomes the representation of the generalized singular eigenfunction

$$\phi_L(z,\mu) = \frac{cz}{2}\frac{g_L^*(z,\mu)}{z-\mu} + \Lambda_L(z)\delta(z-\mu) \qquad (32)$$

for truncated scattering at $L$.

Similarly, the second kind polynomials generate the following generalized singular eigenfunction:

$$\Theta(z,\mu) \equiv \sum_{l=0}^{\infty}\frac{(2l+1)}{2}\rho_l(z)P_l(\mu), \qquad (33)$$

which, from an identical procedure as above, is

$$\Theta_L(z,\mu) = \frac{z}{2}\frac{\left[ch_L^*(z,\mu)-1\right]}{z-\mu} + \gamma_L(z)\delta(z-\mu) \qquad (34)$$

for truncated scattering at $L$ and where

$$h_l^*(z,\mu) \equiv \sum_{j=0}^{l} \omega_j \rho_j(z) P_j(\mu).$$

Note that Eq (31a) reduces to the usual delta function when $z$ is real, and it is single valued on the real axis. In addition, for $z$ not in $[-1,1]$, the LHS of Eq(31a) vanishes. It should also be emphasized that rather than assuming the formal expression for $\delta(z-\mu)$ as above, one can proceed with the finite sum in Eq(26a) and correspondingly for $\Theta_L(z,\mu)$ and take the limit in the last step of the analytical inversion to follow without changing the final result.

Equations (32) and (34) reduce to the Case singular eigenfunctions on the branch cut in the complex $z$-plane as shown below.

**B.2 Evaluation of $T(z,\mu;\mu_0)$**

Consider $\chi_l(z,\mu)$ from Eq(20b) written as

$$\chi_l(z,\mu) \equiv \rho_l(z)\sum_{j=0}^{l}(2j+1)g_j(z)P_j(\mu) - g_l(z)\sum_{j=0}^{l}(2j+1)\rho_j(z)P_j(\mu).$$

and introduce the Darboux relations, Eq(A2.9) and (A2.10), for the partial sums

$$\chi_l(z,\mu) = \frac{z}{z-\mu}\left\{\begin{array}{l}-P_l(\mu)+ \\ +\rho_l(z)cg_l^*(z,\mu) - g_l(z)\left[ch_l^*(z,\mu)-1\right]\end{array}\right\}. \qquad (35)$$

On substitution for $g_l^*(z,\mu)$ and $ch_l^*(z,\mu)-1$ from Eqs(32) and (34),

$$\chi_l(z;\mu) = 2\begin{bmatrix} p_l(z)\phi_l(z,\mu) - g_l(z)\Theta_l(z,\mu) - \\ +zQ_l(z)\delta(z-\mu) - \dfrac{z}{z-\mu}P_l(z) \end{bmatrix}. \tag{36}$$

From the definition Eq(25c), the second term, $T(z,\mu;\mu_0)$, can be shown to be

$$T(z,\mu;\mu_0) = 2S(z,\mu;\mu_0) + \dfrac{z}{z-\mu}\left[\delta(z-\mu) - \delta(\mu-\mu_0)\right], \tag{37a}$$

where

$$S(z,\mu;\mu_0) = \sum_{l=0}^{\infty}\dfrac{(2l+1)}{2}\left[p_l(z)\phi_l(z,\mu_0) - g_l(z)\Theta_l(z,\mu_0)\right]P_l(\mu),$$

and using Eqs(A1.1) and (A1.2) in their limits. The last expression (now with subscript $L$) expressed as

$$S_L(z,\mu;\mu_0) = \left(\sum_{l=0}^{L} + \sum_{l=L+1}^{\infty}\right)\dfrac{(2l+1)}{2}\left[p_l(z)\phi_l(z,\mu_0) - g_l(z)\Theta_l(z,\mu_0)\right]P_l(\mu)$$

becomes, after substitution of the generalized singular eigenfunctions,

$$S_L(z,\mu;\mu_0) = \Theta_L(z,\mu)\phi_L(z,\mu_0) - \Theta_L(z,\mu_0)\phi_L(z,\mu) - \\ -\sum_{l=0}^{L}\dfrac{(2l+1)}{2}\begin{Bmatrix} p_l(z)\left[\phi_l(z,\mu_0) - \phi_L(z,\mu_0)\right] - \\ -g_l(z)\left[\Theta_l(z,\mu_0) - \Theta_L(z,\mu_0)\right] \end{Bmatrix}P_l(\mu). \tag{37b}$$

**B.3 Alternative expression for $\bar{\psi}_{0,L}(k;\mu_0)$ in terms of generalized singular eigenfunctions**

The following symmetry:

$$\bar{\psi}_L(k,\mu;\mu_0) = \bar{\psi}_L(k,\mu_0;\mu) \tag{38}$$

is apparent from Eq(9). Integrating over $\mu$ to give the transform of the scalar flux

$$\bar{\psi}_{0,L}(k;\mu_0) \equiv \int_{-1}^{1} d\mu \bar{\psi}_L(k,\mu;\mu_0) = \int_{-1}^{1} d\mu \bar{\psi}_L(k,\mu_0;\mu)$$

and substituting Eq(25a) with $\mu$ and $\mu_0$ interchanged in the last integral gives

$$\bar{\psi}_{0,L}(k;\mu_0) = \int_{-1}^{1} d\mu \left[ \phi_L(-z,\mu_0) \bar{\psi}_{0,L}(k;\mu) - T(-z,\mu_0;\mu) \right] \tag{39}$$

Since [from Eq(20b)]

$$\int_{-1}^{1} d\mu \chi_l(-z,\mu) = 2\rho_l(-z),$$

we find the individual integrals in Eq(39) to be

$$\int_{-1}^{1} d\mu \bar{\psi}_0(k;\mu) = \frac{1}{\Lambda_L(z)} \left[ \int_{-1}^{1} d\mu \frac{z}{z+\mu} - cz \sum_{l=0}^{L} \omega_l Q_l(-z) \int_{-1}^{1} d\mu \chi_l(-z,\mu) \right] = 2\frac{\gamma_L(z)}{\Lambda_L(z)},$$

where

$$\gamma_L(-z) = \gamma_L(z) \equiv zQ_0(z) - cz \sum_{l=0}^{L} \omega_l Q_l(z) \rho_l(z)$$

and

$$\int_{-1}^{1} d\mu T(-z,\mu_0;\mu) = \sum_{l=0}^{\infty} \frac{(2l+1)}{2} \rho_l(-z) P_l(\mu_0) \equiv \Theta_L(-z,\mu_0)$$

respectively, Therefore, most conveniently

$$\bar{\psi}_{0,L}(k;\mu_0) = 2\phi_L(-z,\mu_0) \frac{\gamma_L(z)}{\Lambda_L(z)} - \Theta_L(-z,\mu_0). \tag{40}$$

## III.C. Fourier transform inversion

By introducing $\bar{\psi}_{0,L}(k;\mu_0)$ from Eq(40) into Eq(25a) with $T(z,\mu;\mu_0)$ from Eqs(37), $\bar{\psi}_L(k,\mu;\mu_0)$ becomes

$$\bar{\psi}_L(k,\mu;\mu_0) = 2\phi_L(-z,\mu)\phi_L(-z,\mu_0)\frac{\gamma_L(z)}{\Lambda_L(z)} - H_L(-z,\mu;\mu_0) \tag{41a}$$

with

$$H_L(z,\mu;\mu_0) \equiv 2\left\{\begin{array}{l} \Theta_L(z,\mu)\phi_L(z,\mu_0) + \\ + \sum_{l=0}^{L}\frac{(2l+1)}{2}\left\{\begin{array}{l}\rho_l(z)\left[\phi_l(z,\mu_0) - \phi_L(z,\mu_0)\right] - \\ -g_l(z)\left[\Theta_l(z,\mu_0) - \Theta_L(z,\mu_0)\right]\end{array}\right\}P_l(\mu) + \\ +\frac{1}{2}\frac{z}{z-\mu}\left[\delta(z-\mu_0) - \delta(\mu-\mu_0)\right]\end{array}\right\}. \tag{41b}$$

Our immediate task is to invert this Fourier transform.

To evaluate the inversion integral given by Eq(8), we consider the following contour integral for $x \geq 0$

$$I_L(x,\mu;\mu_0) \equiv \lim_{\substack{R\to\infty \\ \varepsilon\to 0}} \frac{1}{2\pi}\oint_{C_{R,\varepsilon}} dk\, e^{ikx}\bar{\psi}_L(k,\mu;\mu_0) \tag{42a}$$

closing in the upper half of the complex *k*-plane with

$$C_{R,\varepsilon} \equiv \Gamma_R + C_{R,\varepsilon}^- + \Gamma_{R,\varepsilon}^- + C_\varepsilon + \Gamma_{R,\varepsilon}^+ + C_{R,\varepsilon}^+ \tag{42b}$$

shown in Fig. 1. In the usual way, the evaluation of $I_L(x,\mu;\mu_0)$ will be through the poles and branch cut singularities of the analytically continued integrand. From a reformulation, the desired result is therefore

$$\psi_L(x,\mu;\mu_0) = I_L(x,\mu;\mu_0) - \frac{1}{2\pi}\lim_{\substack{R\to\infty \\ \varepsilon\to 0}} \int_{\substack{C_{R,\varepsilon}^- + \\ +\Gamma_{R,\varepsilon}^- + C_\varepsilon + \\ +\Gamma_{R,\varepsilon}^+ + C_{R,\varepsilon}^+}} dk\, e^{ikx}\bar{\psi}_L(k,\mu;\mu_0). \qquad (43)$$

**C.1 Pole contributions**

The poles come from the denominator of the integrand (the dispersion relation),

$$\Lambda_L(\pm\nu_{0m}) = 0, \quad m=1,\ldots,\mathcal{M}_L, \qquad (44a)$$

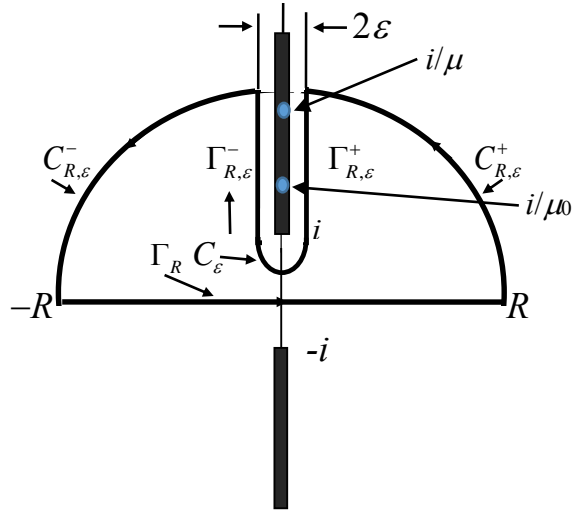

Fig. 1. Complex $k$-plane.

where

$$k_{0m} \equiv -\left(\frac{i}{-\nu_{0m}}\right). \qquad (44b)$$

That the $\mathcal{M}_L$ zeros occur in pairs, are real, simple and of a limited number, has been previously shown (See Inönü [10] and references therein).

From the residue theorem

$$I_L(x,\mu;\mu_0) = i\sum_{m=1}^{\mathcal{M}_L} \mathrm{Res}(\bar{\psi}_L, k_{0m}) e^{ik_{0m}x}. \qquad (45)$$

Let

$$I_m(x,\mu;\mu_0) \equiv Res(\bar{\psi}_L, k_{0m})e^{ik_{0m}x} = \lim_{k \to k_{0m}}\left[(k-k_{0m})\bar{\psi}_L(k,\mu;\mu_0)\right]e^{ik_{0m}x}$$

and transforming to $z$

$$I_m(x,\mu;\mu_0) = i\lim_{z \to v_{0m}}\left[\left(\frac{v-v_{0m}}{v_{0m}z}\right)\left[2\phi_L(z,\mu)\phi_L(z,\mu_0)\frac{\gamma_L(z)}{\Lambda_L(z)} - H(z,\mu;\mu_0)\right]\right]e^{ik_{0m}x}$$

gives

$$I_m(x,\mu;\mu_0) =$$

$$= \frac{i}{v_{0m}^2}\phi_L(v_{0m},\mu)\phi_L(v_{0m},\mu_0)\gamma_L(v_{0m})\lim_{z \to v_{0m}}\left\{2\left[\frac{\Lambda_L(z)}{v-v_{0m}}\right]^{-1} - (v-v_{0m})H(z,\mu;\mu_0)\right\}e^{-x/v_{0m}}.$$

In the limit therefore,

$$I_m(x,\mu;\mu_0) = -\frac{2i}{v_{0m}^2}\phi_L(v_{0m},\mu)\phi_L(v_{0m},\mu_0)\frac{\gamma_L(v_{0m})}{\Lambda'_L(v_{0m})}e^{-x/v_{0m}}. \quad (46)$$

To conform to the standard form of the singular eigenfunction solution [9], we use the expression derived Appendix B

$$\gamma_L(v_{0m}) = \frac{1}{cg_L^*(v_{0m},v_{0m})}\left\{\Lambda_L(v_{0m})\left[ch_L^*(v_{0m},v_{0m})-1\right]+1\right\} \quad (B.5)$$

to give

$$I_m(x,\mu;\mu_0) = -i\frac{\phi_L(v_{0m},\mu)\phi_L(v_{0m},\mu_0)}{M_L(v_{0m})}e^{-x/v_{0m}}$$

with

$$M_L(v_{0m}) \equiv \frac{cv_{0m}^2}{2} g_L^*(v_{0m}, v_{0m}) \Lambda'_L(v_{0m}) \tag{47a}$$

and

$$I_L(x, \mu; \mu_0) = \sum_{m=1}^{M_L} \frac{\phi_L(v_{0m}, \mu)\phi_L(v_{0m}, \mu_0)}{M_L(v_{0m})} e^{-x/v_{0m}}. \tag{47b}$$

It should be noted that Eq(B.5) can also be derived by substitution of Eqs(A2.1a) and (A2.3) into $czg_L^*(z,z)\gamma_L(z)$ and, on simplification, identifying the resulting RHS.

**C.2 Branch cut contribution**
The branch cut arises from the logarithm in the Legendre functions as shown in Eq(A1.4). The integral to evaluate therefore is on the contours $\Gamma_{\infty,\varepsilon}^- + \Gamma_{\infty,\varepsilon}^+$ with $k_\pm \equiv \pm\varepsilon + iu$, $u \in [1,\infty)$. For the change of variable $z_\pm = (ik_\pm)^{-1}$, the branch cut contribution to $I_L(x, \mu; \mu_0)$ becomes

$$I_{\Gamma^++\Gamma^-}(x, \mu; \mu_0) = \frac{i}{2\pi} \lim_{\varepsilon \to 0} \left[ \begin{array}{l} \int_{-i\varepsilon}^{-1-i\varepsilon} dz_+ \frac{e^{x/z_+}}{z_+^2} \overline{\psi}_L(k(z_+), \mu; \mu_0) + \\ + \int_{i\varepsilon}^{-1+i\varepsilon} dz_+ \frac{e^{x/z_-}}{z_-^2} \overline{\psi}_L(k(z_-), \mu; \mu_0) \end{array} \right] \tag{48}$$

over the contour shown in Fig. 2. Continuing with the substitution

$$z_\pm = -v \mp i\varepsilon + O(\varepsilon^2), \quad v \in [0,1],$$

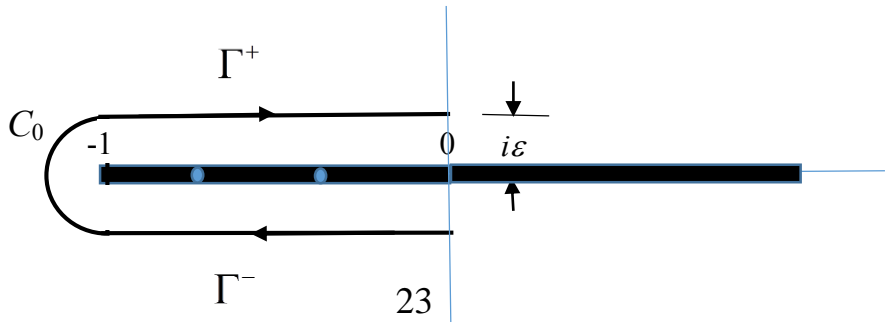

Fig. 2. Path along the branch cut in the $z$-plane.

then

$$I_{L,\Gamma^{+}+\Gamma^{-}}(x,\mu;\mu_0) = -\frac{i}{2\pi}\int_0^1 dv \frac{e^{-x/v}}{v^2} Disc\left[\bar{\psi}_L(-k(v),\mu;\mu_0)\right], \quad (49a)$$

where $v^2\varepsilon$ has been replaced by $\varepsilon$, and

$$Disc\left[\bar{\psi}_L(-k(v),\mu;\mu_0)\right] \equiv$$
$$\equiv \bar{\psi}_L^+(-k(v),\mu;\mu_0) - \bar{\psi}_L^-(-k(v),\mu;\mu_0). \quad (49b)$$

The boundary values of $\bar{\psi}_L$ are

$$\bar{\psi}_L^\pm(-k(v),\mu;\mu_0) \equiv \lim_{\varepsilon\to 0}\bar{\psi}_L(-k(v\pm i\varepsilon),\mu;\mu_0). \quad (50b)$$

Determination of all boundary values begins with the Plemelj formula [9]

$$\left[\frac{1}{z-\mu}\right]^\pm \equiv \lim_{\varepsilon\to 0}\left[\frac{1}{v-(\mu\mp i\varepsilon)}\right] = \boldsymbol{P}\frac{1}{v-\mu}\mp i\pi\delta(v-\mu),$$

where $\boldsymbol{P}$ is the principal value of any integral over the expression following. Hence, the following boundary values result:

a. Legendre functions

$$Q_l^\pm(z) = \frac{1}{2}\int_{-1}^1 d\mu P_l(\mu)\left[\boldsymbol{P}\frac{1}{v-\mu}\mp i\pi\delta(v-\mu)\right]$$
$$= Q_l^*(v)\mp \frac{i\pi}{2}P_l(v), \quad (50a)$$

where

$$Q_l^*(v) = \frac{1}{2} P \int_{-1}^{1} d\mu \frac{P_l(\mu)}{v - \mu}$$

b. Dispersion relation

$$\Lambda_l^{\pm}(z) = (l+1)\left[g_{l+1}(z)Q_l^{\pm}(z) - g_l(z)Q_{l+1}^{\pm}(z)\right]$$

$$\Lambda_l^{\pm}(z) = \Lambda_l^*(v) \pm \frac{i\pi}{2} cv g_l^*(v,v) \tag{50b}$$

c. Axillary relation

$$\gamma_l^{\pm}(z) = \gamma_l^*(v) \pm \frac{i\pi v}{2}\left[ch_l^*(v,v) - 1\right] \tag{50c}$$

d. Generalized singular eigenfunctions

$$\phi_L^{\pm}(z,\mu) = \frac{cv}{2} g_L^*(v,\mu)\left[\frac{1}{z-\mu}\right]^{\pm} + \Lambda_L^{\pm}(z)\delta(v-\mu)$$

$$\phi_L^{\pm}(z,\mu) = \frac{cv}{2} g_L^*(v,\mu)\left[P\frac{1}{v-\mu} \mp i\pi\delta(v-\mu)\right] +$$

$$+ \left[\Lambda_L^*(v) \mp \frac{i\pi}{2} cv g_L^*(v,v)\right]\delta(v-\mu)$$

to give

$$\phi_L^{\pm}(z,\mu) = \frac{cv}{2} P \frac{g_L^*(v,\mu)}{v-\mu} + \Lambda_L^*(v)\delta(v-\mu)$$

$$\equiv \phi_L(v,\mu) \tag{50d}$$

and similarly,

$$\Theta_L^{\pm}(z,\mu) = \frac{\nu}{2} P \frac{\left[ch_i^*(\nu,\mu)-1\right]}{\nu-\mu} + \gamma_L^*(\nu)\delta(\nu-\mu) \tag{50e}$$

$$\equiv \Theta_L(\nu,\mu)$$

Note that $\phi_L^{\pm}(z,\mu)$ [and $\Theta_L^{\pm}(z,\mu)$] become the common singular eigenfunction from their generalization on the branch cut.

With the above boundary values, one finds

$$\overline{\psi}_L^{\pm}(-k(\nu),\mu;\mu_0) \equiv 2\phi_L^{\pm}(z,\mu)\phi_L^{\pm}(z,\mu_0)\frac{\gamma_L^{\pm}(z)}{\Lambda_L^{\pm}(z)} - H_L^{\pm}(z,\mu;\mu_0)$$

and therefore

$$Disc\left[\overline{\psi}(-k(\nu),\mu;\mu_0)\right] =$$
$$= 2\phi_L(\nu,\mu)\phi_L(\nu,\mu_0)\left[\frac{\gamma_L^+(z)}{\Lambda_L^+(z)} - \frac{\gamma_L^-(z)}{\Lambda_L^-(z)}\right] - Disc\left[H_L(\nu,\mu;\mu_0)\right]; \tag{51a}$$

but

$$Disc\left[H_L(z,\mu;\mu_0)\right] = Disc\left[\frac{\delta(z-\mu_0)-\delta(\mu-\mu_0)}{z-\mu}\right]$$

$$= \left\{ \begin{bmatrix} \frac{1}{z-\mu} \end{bmatrix}^+ \lim_{\varepsilon \to 0}\left[\delta(\nu+i\varepsilon-\mu_0)-\delta(\mu-\mu_0)\right] - \right.$$
$$\left. -\begin{bmatrix} \frac{1}{z-\mu} \end{bmatrix}^- \lim_{\varepsilon \to 0}\left[\delta(\nu-i\varepsilon-\mu_0)-\delta(\mu-\mu_0)\right] \right\}$$

giving

$$Disc\left[H_L(z,\mu;\mu_0)\right] = \left\{\begin{bmatrix} \frac{1}{z-\mu} \end{bmatrix}^+ - \begin{bmatrix} \frac{1}{z-\mu} \end{bmatrix}^-\right\}\left[\delta(\nu-\mu_0)-\delta(\mu-\mu_0)\right] \tag{51b}$$
$$= 2i\pi\delta(\nu-\mu)\left[\delta(\nu-\mu_0)-\delta(\mu-\mu_0)\right] = 0.$$

From Eq(B.6) of Appendix B

$$\frac{\gamma_L(z)}{\Lambda_L(z)} = \frac{1}{cg_L^*(z,z)}\left\{\left[ch_L^*(z,z)-1\right]+\frac{1}{\Lambda_L(z)}\right\}, \qquad (B.6)$$

the boundary values

$$\frac{\gamma_L^{\pm}(z)}{\Lambda_L^{\pm}(z)} = \frac{1}{cg_L^*(v,v)}\left\{\left[ch_L^*(v,v)-1\right]+\frac{1}{\Lambda_L^{\pm}(z)}\right\}.$$

give across the cut

$$\frac{\gamma_L^+(z)}{\Lambda_L^+(z)} - \frac{\gamma_L^-(z)}{\Lambda_L^-(z)} = \frac{1}{cg_L^*(v,v)}\left[\frac{1}{\Lambda_L^+(z)}-\frac{1}{\Lambda_L^-(z)}\right] = -\frac{\pi i v}{\Lambda_L^+(z)\Lambda_L^-(z)}.$$

Finally, from Eqs(49b) and Eqs(50d,e)

$$Disc\left[\overline{\psi}\left(-k(v),\mu;\mu_0\right)\right] = -2\pi i v\frac{\phi_L(v,\mu)\phi_L(v,\mu_0)}{\Lambda_L^+(v)\Lambda_L^-(v)}. \qquad (52)$$

Therefore, the branch cut contribution from Eqs(49a) is

$$I_{L,\Gamma^++\Gamma^-}(x,\mu;\mu_0) = -\int_0^1 dv\frac{e^{-x/v}}{M_L(v)}\phi_L(v,\mu)\phi_L(v,\mu_0) \qquad (53a)$$

with

$$M_L(v) \equiv v\Lambda_L^+(v)\Lambda_L^-(v). \qquad (53b)$$

**C.3 Assembling the final solution**

Noting that the contributions from the contours $C_\varepsilon$ and $C_{R,\varepsilon}^{\pm}$ vanish in their respective limits, Eqs(43), (47) and Eq(53) give the final solution for truncated scattering for $x \geq 0$ as

$$\psi_L(x,\mu;\mu_0) = \sum_{m=1}^{M_L} \frac{\phi_L(\nu_{0m},\mu)\phi_L(\nu_{0m},\mu_0)}{M_L(\nu_{0m})} e^{-x/\nu_{0m}} +$$
$$+ \int_0^1 dv \frac{e^{-x/\nu}}{M_L(\nu)} \phi_L(\nu,\mu)\phi_L(\nu,\mu_0). \quad (54a)$$

Since from reciprocity

$$\psi_L(-|x|,\mu;\mu_0) = \psi_L(|x|,-\mu;-\mu_0)$$

and $\phi_L(-\nu,\mu) = \phi_L(\nu,-\mu)$, Eq(54a) also gives

$$\psi_L(-|x|,\mu;\mu_0) = \sum_{m=1}^{M_L} \frac{\phi_L(-\nu_{0m},\mu)\phi_L(-\nu_{0m},\mu_0)}{M_L(\nu_{0m})} e^{-|x|/\nu_{0m}} +$$
$$+ \int_0^1 dv \frac{e^{-|x|/\nu}}{M_L(\nu)} \phi_L(-\nu,\mu)\phi_L(-\nu,\mu_0). \quad (54b)$$

In the limit of no truncation, $L \to \infty$, therefore, we arrive at the classical singular eigenfunction expansion for $x \gtrless 0$

$$\psi(x,\mu;\mu_0) = \sum_{m=1}^{M_L} \frac{\phi(\pm\nu_{0m},\mu)\phi(\pm\nu_{0m},\mu_0)}{M(\nu_{0m})} e^{-|x|/\nu_{0m}} +$$
$$+ \int_0^1 dv \frac{e^{-|x|/\nu}}{M(\nu)} \phi(\pm\nu,\mu)\phi(\pm\nu,\mu_0), \quad (55a)$$

where

$$\phi(\nu,\mu) \equiv \lim_{L\to\infty}\phi_L(\nu,\mu) = \frac{c\nu}{2}\mathbf{P}\frac{g^*(\nu,\mu)}{\nu-\mu} + \Lambda^*(\nu)\delta(\nu-\mu) \quad (55b)$$

with

$$g^*(\nu,\mu) \equiv \sum_{l=0}^{\infty} \omega_l g_l(\nu) P_l(\mu) \quad (55c)$$

$$\Lambda^*(v) \equiv 1 - cv \sum_{l=0}^{\infty} \omega_l Q_l(v) g_l(v). \tag{55d}$$

**FINAL REMARKS**
The above derivation merits additional comment. While the result is not new, the steps getting to it are. The primary reason the approach succeeds is that it is a consequence of the solution to the moments recurrence coming from an analytical closure. This is novel since an analytical solution to a recurrence is not common in solutions to the transport equation. Recurrences generally find use in numerical, not theoretical evaluations. It is also obvious that the approach works because we know what we are to look for—the singular eigenfunction expansion. Thus, it is not too surprising that, without the guidance of Case, the singular eigenfunction solution was not discovered from Fourier transforms. Additionally, no orthogonality or completeness is evident as the eigenfunction expansion simply emerges from manipulation in the complex plane. For this reason, the Fourier transform derivation is certainly less elegant than Case's solution.

Essentially, the singular eigenfunction expansion becomes clear through organization of the transform into singular and analytical components, facilitated by identification of the generalized singular eigenfunctions. This enables the first term of the transform in Eq(41a) to bear the singularities; while, the second to be analytic. Note that all singularities originate from the dispersion relation. It is also clear that *two* generalized eigenfunctions are involved; while, only *one* participates in the final result. That there are two comes from the two independent solutions to the moments recurrence. Even though, the additional generalized singular eigenfunction does not participate, it greatly facilitates the derivation, which apparently is its role.

To the author's knowledge, the generalized singular eigenfunctions, whose existence, based on the Legendre polynomial expansion for a generalized delta function, are new. The final result, however, does not depend upon this definition as shown by the less intuitive derivation in Ref. 6. There, no mention of generalized eigenfunctions is made. The benefit of generalized functions apparently comes from the organization they provide in the Fourier transform space. It is also clear that we have tacitly assumed analyticity of the generalized delta function. Most likely, this is justified by requiring the delta functional to operate on a suitably smooth function space to guarantee its analytic properties. Also as mentioned, one can avoid the definition of the delta function entirely by enforcing the limit in $N$ as the last step in the determination of Eq(54a).

Finally, it is interesting to note that the assumption of truncating the scattering kernel, which is usually out of practical necessity, is purely artificial to enable the matrix solution for the moments. It has no influence on the singular eigenfunction expansion. as long as the condition of Eq(1c) is satisfied.

**Appendix A: Reference Formulas**
The formulas involving Legendre polynomials and functions separately and mixed with Chandrasekhar polynomials are assembled here for convenience.

**A1. Legendre polynomials/functions**
For $L$ a non-negative integer:

$$\sum_{l=0}^{L} \frac{(2l+1)}{2} P_l(z) P_l(\mu) = \frac{L+1}{2} \left[ \frac{P_{L+1}(z) P_L(\mu) - P_L(z) P_{L+1}(\mu)}{z - \mu} \right] \quad (A1.1)$$

$$\sum_{l=0}^{L} \frac{(2l+1)}{2} Q_l(z) P_l(\mu) = \frac{1}{2} \frac{1}{z - \mu} - \frac{L+1}{2} \left[ \frac{P_{L+1}(\mu) Q_L(z) - P_L(\mu) Q_{L+1}(z)}{z - \mu} \right] \quad (A1.2)$$

$$1 = (L+1)\left[ P_{L+1}(z) Q_L(z) - P_L(z) Q_{L+1}(z) \right] \quad (A1.3)$$

$$Q_L(z) = P_L(z) \frac{1}{2} \ln\left[\frac{z+1}{z-1}\right] - W_{L-1} \quad (A1.4)$$

$W_L$ is the Legendre polynomial of the second kind.

**A2. Legendre polynomials/functions mixed with Chandrasekhar polynomials**

$$cz g_l^*(z,z) = (l+1)\left[ g_l(z) P_{l+1}(z) - g_{l+1}(z) P_l(z) \right] \quad (A2.1a)$$

where

$$g_l^*(z,\mu) \equiv \sum_{j=0}^{l} \omega_j g_j(z) P_j(\mu) \quad \text{(A2.1b)}$$

$$z\left[ch_l^*(z,z)-1\right] = (l+1)\left[\rho_l(z)P_{l+1}(z)-\rho_{l+1}(z)P_l(z)\right] \quad \text{(A2.2a)}$$

where

$$h_l^*(z,\mu) \equiv \sum_{j=0}^{l} \omega_j \rho_j(z) P_j(\mu) \quad \text{(A2.2b)}$$

$$\gamma_l(z) \equiv (l+1)\left[\rho_{l+1}(z)Q_l(z)-\rho_l(z)Q_{l+1}(z)\right] \quad \text{(A2.3)}$$

$$\Lambda_l(z) = (l+1)\left[g_{l+1}(z)Q_l(z)-g_l(z)Q_{l+1}(z)\right] \quad \text{(A2.4)}$$

$$\psi_l(z) = czg_l^*(z,z) = (l+1)\left[g_l(z)P_{l+1}(z)-g_{l+1}(z)P_l(z)\right] \quad \text{(A2.5)}$$

$$\theta_l(z) = z\left[ch_l^*(z,z)-1\right] = (l+1)\left[\rho_l(z)P_{l+1}(z)-\rho_{l+1}(z)P_l(z)\right] \quad \text{(A2.6)}$$

$$z = l\left[\rho_l(z)g_{l-1}(z)-g_l(z)\rho_{l-1}(z)\right] \quad \text{(A2.7)}$$

$$g_l(z)\gamma_l(z) - \rho_l(z)\Lambda_l(z) = zQ_l(z) \quad \text{(A2.8)}$$

$$\sum_{j=0}^{l}(2j+1)g_j(z)P_j(\mu) =$$

$$= \frac{1}{z-\mu}\left\{\begin{array}{c}(l+1)\left[g_{l+1}(z)P_l(\mu)-g_l(z)P_{l+1}(\mu)\right]+\\ +czg_l^*(z,\mu)\end{array}\right\} \quad \text{(A2.9)}$$

$$\sum_{j=0}^{l}(2j+1)\rho_j(z)P_j(\mu)=$$
$$=\frac{1}{z-\mu}\left\{\begin{array}{c}(l+1)\left[\rho_{l+1}(z)P_l(\mu)-\rho_l(z)P_{l+1}(\mu)\right]+\\+z\left[ch_l^*(z,\mu)-1\right]\end{array}\right\} \quad (A2.10)$$

**Appendix B: An alternative expression for** $\dfrac{\gamma_L(z)}{\Lambda_L(z)}$

From Eqs(A2.3) and (A2.7), we first find an expression for $\rho_L(z)$ from Cramer's rule

$$\rho_L(z)=\frac{\begin{vmatrix}\gamma_L(z)/(L+1) & Q_L(z)\\ z/(L+1) & g_L(z)\end{vmatrix}}{\begin{vmatrix}-Q_{L+1}(z) & Q_L(z)\\ -g_{L+1}(z) & g_L(z)\end{vmatrix}}$$

or

$$\rho_L(z)=\frac{g_L(z)\gamma_L(z)-zQ_L(z)}{-\Lambda_L(z)} \quad (B.1)$$

Next, find a second expression for $\rho_L(z)$ from Eq(A2.2a) and (A2.3)

$$\rho_L(z)=\frac{\begin{vmatrix}z\left[ch_L^*(z,z)-1\right]/(L+1) & -P_L(z)\\ \gamma_L(z)/(L+1) & Q_L(z)\end{vmatrix}}{\begin{vmatrix}P_{L+1}(z) & -P_L(z)\\ Q_{L+1}(z) & -Q_L(z)\end{vmatrix}}$$

$$\rho_L(z)=-\left\{z\left[ch_L^*(z,z)-1\right]Q_L(z)+\gamma_L(z)P_L(z)\right\} \quad (B.2)$$

Equating Eqs(B.1) and (B.2) gives

$$\gamma_L(z)\left[g_L(z)-P_L(z)\Lambda_L(z)\right]=z\left\{\Lambda_L(z)\left[ch_L^*(z,z)-1\right]+1\right\}Q_L(z) \quad (B.3)$$

Finally, find $g_L(z)$ from Eqs(A2.1a) and (A2.4)

$$g_L(z)=\frac{\begin{vmatrix}czg_L^*(z,z)/(L+1) & -P_L(z)\\ \Lambda_L(z)/(L+1) & Q_L(z)\end{vmatrix}}{\begin{vmatrix}P_{L+1}(z) & -P_L(z)\\ Q_{L+1}(z) & -Q_L(z)\end{vmatrix}}$$

or

$$g_L(z)=czg_L^*(z,z)Q_L(z)+P_L(z)\Lambda_L(z), \quad (B.4)$$

and substitute into Eq(B.3) to give

$$\gamma_L(z)=\frac{1}{cg_L^*(z,z)}\left[1+\Lambda_L(z)\left[ch_L^*(z,z)-1\right]\right] \quad (B.5)$$

and the desired result

$$\frac{\gamma_L(z)}{\Lambda_L(z)}=\frac{1}{cg_L^*(z,z)}\left[\frac{1}{\Lambda_L(z)}+ch_L^*(z,z)-1\right]. \quad (B.6)$$